\documentclass[journal=apchd5,manuscript=article,layout=twocolumn]{achemso}

\usepackage{chemformula} % Formula subscripts using \ch{}

\usepackage{amsmath,amssymb,pifont}
\usepackage{color}
\usepackage{graphicx}
\usepackage{pstool}
\usepackage{makecell}
\usepackage{amsfonts} 
\usepackage{dblfloatfix} 
\usepackage{threeparttable}
\usepackage{amsfonts}
\usepackage{multirow}
\usepackage{float}
\usepackage{graphicx,bm}
\usepackage{subfig}
\usepackage{cancel}
\usepackage{pstool,pstricks}

\usepackage[thinlines]{easytable}

\newcommand{\ve}[1]{\mathbf{#1}}
\newcommand{\te}[1]{\overline{\overline{\mathbf{#1}}}}

\author{Karim Achouri}
\email{karim.achouri@epfl.ch}
\author{Mintae Chung}
\author{Andrei Kiselev}
\author{Olivier J. F. Martin}

\affiliation[EPFL]
{Nanophotonics and Metrology Laboratory, Institute of Electrical and Microengineering, \'Ecole Polytechnique F\'ed\'erale de Lausanne, Route Cantonale, 1015 Lausanne, Switzerland.}

\title{Multipolar Pseudochirality\\ Induced Optical Torque}
\abbreviations{IR,NMR,UV}

\begin{document}

\maketitle
%\tableofcontents

\begin{abstract}
It has been observed that achiral nano-particles, such as flat helices, may be subjected to an optical torque even when illuminated by normally incident linearly polarized light. However, the origin of this fascinating phenomenon has so far remained mostly unexplained. We therefore propose an exhaustive discussion that provides a clear and rigorous explanation for the existence of such a torque. Using multipolar theory, and taking into account nonlocal interactions, we find that this torque stems from multipolar pseudochiral responses that generate both spin and orbital angular momenta. We also show that the nature of these peculiar responses makes them particularly dependent on the asymmetry of the particles. By elucidating the origin of this type of torque, this work may prove instrumental for the design of high-performance nano-rotors.
\end{abstract}

\section{Introduction}

The ability to optically manipulate microscopic particles has been a tremendous source of interest since the first demonstration of optical tweezers back in 1986~\cite{ashkin1986,grier2003}. Since then, we have witnessed the emergence of a plethora of related concepts and applications that, for instance, apply to biosensing~\cite{lafleur2016}, drug delivery~\cite{guix2014}, cell trapping~\cite{gultepe2013} and nano-mechanical motion control using nano-motors~\cite{yan2012,kim2016a,shao2018}. In more recent years, the capabilities to control optical forces and torques was further improved with the development of metamaterials, which exhibit unprecedented degrees of freedom to engineer the propagation of light~\cite{tang2020,kotsifaki2021,achouri2021,shi2022a}.

In this work, we are specifically interested in the development of optically-driven nano-rotors that are able to convert the momentum of light into mechanical rotation~\cite{shao2018}. This is typically achieved in a particle by absorption or scattering of light possessing spin angular momentum (SAM)~\cite{beth1936} or orbital angular momentum (OAM).~\cite{allen1992,yao2011,padgett2017} Many examples of such optical torque manipulations may be found in the literature for both spin~\cite{friese1998,leach2006,yan2012,lehmuskero2013} and orbital~\cite{ladavac2004,loke2014,shen2016,parker2020} types of angular momenta.

The ability to transfer angular momentum to a particle can generally be assessed by considering its effective material parameters. In the simplest scenario, isotropic particles may be subjected to an optical rotation due to their ability to scatter or absorb energy from an illumination \emph{already} possessing SAM and/or OAM. In this case, the transfer of angular momentum is a process that fully depends on the illumination condition and that can be easily assessed from the electric polarizability of the particles~\cite{chaumet2009,vesperinas2010, vesperinas2015}. A more advanced control of the optical torque may be achieved by using particles with birefringent, anisotropic or chiral properties that react differently for illuminations with opposite circular polarization handedness or, alternatively, that induce rotation of polarization or conversion for linearly polarized excitations~\cite{kim2016a,shen2016, shao2018,parker2020}.
\begin{figure*}[t!]
	\centering
	\includegraphics[width=1.75\columnwidth]{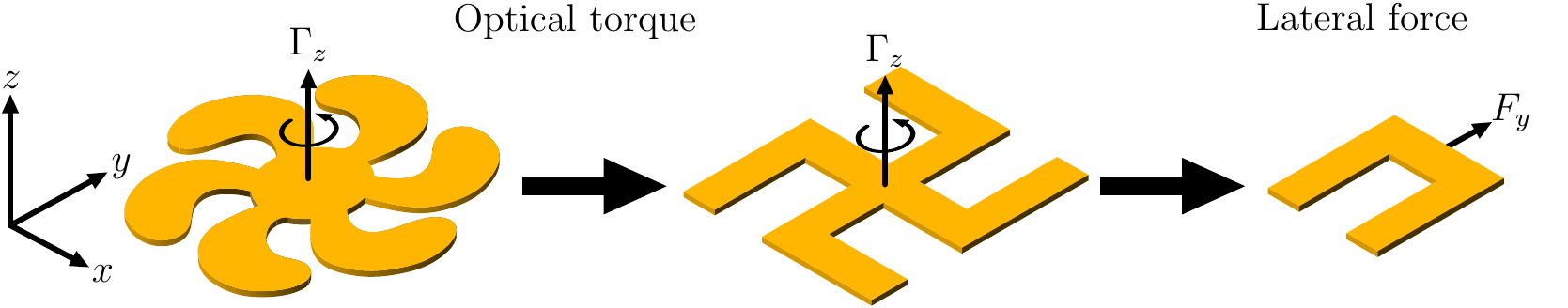}
	%		\psfragfig*[width=2\columnwidth]{figs/concept}{
	%			\psfrag{a}[][][1.]{Optical torque}
	%			\psfrag{b}[][][1.]{Lateral force}
	%			\psfrag{t}[][][1.]{$\Gamma_z$}
	%			\psfrag{f}[][][1.]{$F_y$}
	%			\psfrag{x}[][][1.]{$x$}
	%			\psfrag{y}[][][1.]{$y$}
	%			\psfrag{z}[][][1.]{$z$}}
	\caption{Problem simplification for different particles subjected to optical forces and torques when illuminated by a $z$-propagating plane wave.}
	\label{fig_concept}
\end{figure*}

This work concentrates on particles exhibiting chiral responses (optical activity) as they may experience an optical torque under a linearly polarized excitation with an arbitrary orientation of the polarization axis, which is particularly attractive in practical applications. To generate a chiral response, one may a priori think that it is required to consider particles that are 3D geometrically chiral meaning that they cannot be superimposed with their mirror image by combination of rotations and translation operations. However, this type of chirality, sometimes referred to as \emph{intrinsic} chirality, is not necessarily required to achieve a chiral response. Indeed, geometrically achiral structures may still exhibit chiral responses provided that the superposition of the illumination with the structure itself leads to an overall arrangement that has no plane, center or axis of symmetry~\cite{arnaut1997,plum2007,plum2009,okamoto2019}. In this case, this type of chirality is commonly referred to as \emph{extrinsic} chirality or pseudochirality as it does not only depend on the structure itself but also on the properties of an external illumination, such as illuminating a 2D array of achiral particles at \emph{oblique} incidence~\cite{arnaut1997,tretyakov2001,plum2007}. 

Importantly, these concepts do not only apply to 3D objects but also to 2D structures that can exhibit intrinsic/extrinsic planar chirality~\cite{hecht1994,papakostas2003,plum2007,plum2009,okamoto2019}. As such, flat spiral-shaped particles with threefold (or more) rotational symmetry are neither birefringent nor 3D chiral~\cite{achouri2021a}, but can nonetheless exhibit different polarization effects when illuminated at \emph{normal} incidence in opposite direction with respect to their plane of symmetry~\cite{hecht1994,papakostas2003}. However, these effects have been reported to be extremely weak, if not completely nonexistent, in the case of optically thin plasmonic helices unless a substrate is present to break the symmetry of the system along the illumination direction~\cite{gonokami2005,arteaga2016,okamoto2019}. The reason for the reported nonexistence of optical activity in these cases, which may appear as a paradox compared to the previously cited Refs.~\cite{hecht1994,papakostas2003}, stems from the fact that the considerations in Refs.~\cite{gonokami2005,arteaga2016,okamoto2019} are restricted to the dipolar regime. If higher-order multipolar and nonlocal responses are taken into account, then the existence of pseudochiral responses from 2D chiral structures may be explained even for normally propagating waves, as recently demonstrated in Ref.~\cite{achouri2022}

In the context of optical torque, this begs the question as to what is the fundamental origin of the torque experienced by flat spiral-shaped particles embedded in a homogeneous medium when illuminated by a linearly polarized plane wave at \emph{normal} incidence, as reported in Refs~\cite{liu2010b,chung2022}. Currently, the only explanation is that the light scattered near the arms of such particles possesses OAM.~\cite{liu2010b} However, this explanation remains unsatisfactory as it does not define the origin and the mechanisms that generate such momentum. Additionally, it remains unclear whether pseudochiral effects play a role and if SAM is really nonexistent in this process, as suggested in Ref.~\cite{liu2010b} In principle, it should be possible to predict the existence of such a torque directly from the particle effective material parameters, which could themselves be deduced by considering the interactions between the multipoles excited in the particle~\cite{chen2011a,bliokh2015,achouri2020}. However, the investigation of the interactions between various multipolar contributions and their relation with the emergence of an optical torque has not yet been applied to the current situation.

The purpose of this work is thus to provide an extensive explanation of the underlying mechanisms responsible for the optical torque acting on geometrically achiral nano-particles using multipolar theory. Note that we shall not directly investigate the forces and torques acting on intricate helical shaped nano-structures due to the excessive complexity of their electromagnetic responses. Instead, and without loss of generality, we simplify the problem by investigating the optical torque acting on a Gammadion particle, as depicted in Fig.~\ref{fig_concept}. To provide an initial intuitive understanding of the origin of such a torque, we shall start our discussion by considering the lateral force acting on a U-shaped section of this Gammadion particle. By using the theory developed in Ref.~\cite{achouri2022}, we show in Sec.~2 that a simple U-shape structure develops an extraordinary magnetic dipolar response parallel to the direction of wave propagation leading to an asymmetric scattering response responsible for a lateral force. In Sec.~3, we use this effect to qualitatively explain the origin of the torque on the Gammadion particle in terms of the lateral forces acting on its U-shaped blades. In Sec.~4, we use full-wave simulations to analyze the multipolar responses of the Gammadion particle and show that they are indeed responsible for the origin of the observed optical torque. To cast further light on the origin of these multipoles, we also inspect their emergence in terms of the asymmetry of the Gammadion particle as its structure changes from a regular symmetric cross to the Gammadion shape depicted in Fig.~\ref{fig_concept}. Finally, we conclude in Sec.~5.

\section{Asymmetric Optical Force}
\label{sec_AOF}

\begin{figure*}[t!]
	\centering	\includegraphics[width=2\columnwidth]{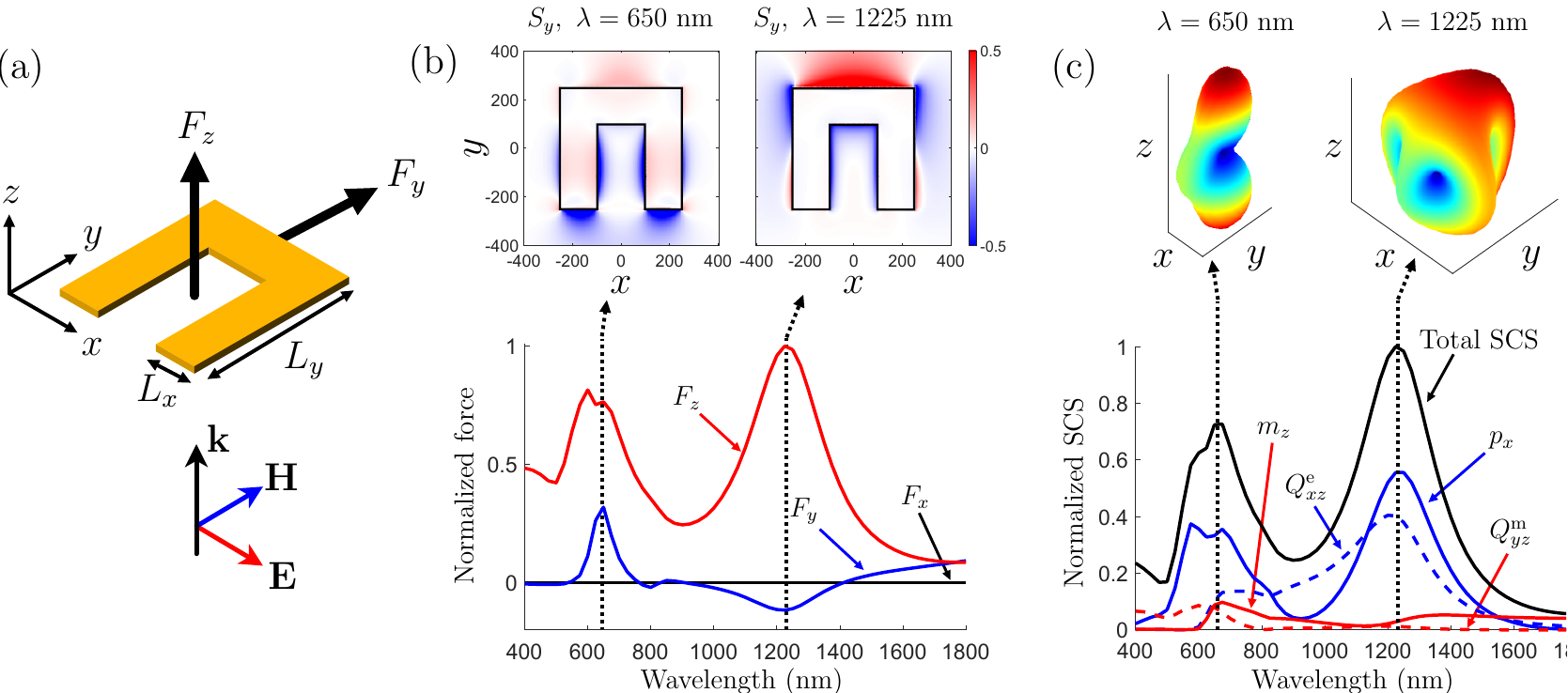}
%			\psfragfig*[width=2\columnwidth]{figs/U-shape}{
%			\psfrag{a}[][][1.]{(a)}
%			\psfrag{b}[][][1.]{(b)}
%			\psfrag{c}[][][1.]{(c)}
%			\psfrag{lx}[][][1.]{$L_x$}
%			\psfrag{ly}[][][1.]{$L_y$}
%			\psfrag{aa}[][][0.7]{$p_x$}
%			\psfrag{bb}[][][0.7]{$m_z$}
%			\psfrag{cc}[][][0.7]{$Q^\text{e}_{xz}$}
%			\psfrag{dd}[][][0.7]{$Q^\text{m}_{yz}$}
%			\psfrag{mm}[][][0.7]{$S_y,~\lambda=650$~nm}
%			\psfrag{ii}[][][0.7]{$S_y,~\lambda=1225$~nm}
%			\psfrag{ee}[][][0.7]{Total SCS}
%			\psfrag{ff}[][][0.7]{$F_z$}
%			\psfrag{hh}[][][0.7]{$F_y$}
%			\psfrag{gg}[][][0.7]{$F_x$}
%			\psfrag{p}[][][0.7]{Normalized force}
%			\psfrag{i}[][][0.7]{Wavelength (nm)}
%			\psfrag{j}[][][0.7]{Normalized SCS}
%			\psfrag{g}[][][0.7]{$\lambda=1225$~nm}
%			\psfrag{f}[][][0.7]{$\lambda=650$~nm}
%			\psfrag{k}[][][1.]{$\ve{k}$}
%			\psfrag{e}[][][1.]{$\ve{E}$}
%			\psfrag{h}[][][1.]{$\ve{H}$}
%			\psfrag{m}[][][1.]{$F_z$}
%			\psfrag{n}[][][1.]{$F_y$}
%			\psfrag{xx}[][][0.7]{$x (\text{nm})$}
%			\psfrag{yy}[][][0.7]{$y (\text{nm})$}
%			\psfrag{x}[][][1.]{$x$}
%			\psfrag{y}[][][1.]{$y$}
%			\psfrag{z}[][][1.]{$z$}}
	\caption{Lateral force acting on an asymmetric gold particle. (a)~Illumination condition. (b)~Components of the optical force and Poynting vector $S_y=\text{Re}\{E_zH_x^*-E_x^*H_z\}/2$. (c) Multipolar decomposition given in terms of the main components of the electric dipole $\ve{p}$, the magnetic dipole $\ve{m}$, the electric quadrupole $\te{Q}^\text{e}$ and the magnetic quadrupole $\te{Q}^\text{m}$. The radiation patterns at the two resonant frequencies are also plotted at the top of (c). The particle dimensions are $L_x=150$~nm, $L_y=500$~nm and a thickness of 30~nm.}
	\label{fig_u_shape}
\end{figure*}

Consider the simple U-shaped optical resonator made of gold that is depicted in Fig.~\ref{fig_u_shape}a and that may be considered as a building block of the Gammadion particle in Fig.~\ref{fig_concept}. This resonator is surrounded by vacuum and, when excited by an $x$-polarized $z$-propagating planewave, as shown in Fig.~\ref{fig_u_shape}a, it experiences a longitudinal radiation pressure force, $F_z$, and a lateral force, $F_y$. Using the Maxwell stress tensor (see the Appendix), these forces are calculated and plotted in Fig.~\ref{fig_u_shape}b. 

While the existence of the force $F_z$ may be intuitively understood in terms of radiation pressure, i.e., the light ``pushes'' on the particle making it move forward, the presence of the lateral force $F_y$ is more difficult to explain. Moreover, note that while the longitudinal radiation pressure force is always positive as function of the wavelength, that is not the case for the lateral force, which is positive at the first resonant frequency ($\lambda=650$~nm) and negative at the second one ($\lambda=1225$~nm). Note that a similar flip of the lateral force as function of plasmonic resonances has already been observed in Refs.~\cite{raziman2015,achouri2020}

The reason that explains the existence of this lateral force, and its strong frequency dependence, is the asymmetry of the particle~\cite{albooyeh2016a,magallanes2018,tanaka2020}. Indeed, the illuminating plane wave excites in the particle a combination of multipoles that results in an overall asymmetric radiation pattern, as imposed by the broken reflection symmetry of the particle along the $y$-axis. This is clearly visible in the multipolar decomposition plotted in Fig.~\ref{fig_u_shape}c that shows the total scattering cross-section (SCS) of the particle and the main contributions from electric and magnetic dipolar and quadrupolar responses. Also plotted in the figure are the total radiation patterns at the two resonant frequencies, which are clearly asymmetric in the $y$-direction as expected from the asymmetry of the particle.

To understand why this superposition of multipoles leads to a lateral force, we now formulate a simplified model of the scattering particle in Fig.~\ref{fig_u_shape}a. For simplicity, we only consider its dipolar contributions whose resulting electric far-field  may be expressed as~\cite{jackson1999}
\begin{equation}
	\label{eq_dip}
	\ve{E} = \frac{e^{-jkr}}{4\pi r}\frac{k^2}{\epsilon_0}\left[\left(\bm{\hat{\theta}}+\bm{\hat{\phi}}\right)\cdot\ve{p} + \frac{1}{c_0} \ve{m}\times\ve{\hat{r}} \right],
\end{equation}
where $\ve{p}=A_\textrm{p}e^{j\varphi_\text{p}}\ve{\hat{x}}$ and $\ve{m}=A_\textrm{m}e^{j\varphi_\text{m}}\ve{\hat{z}}$ with $A_\textrm{p/m}$ and $\varphi_\textrm{p/m}$ being the magnitude and phase of the electric and magnetic dipoles excited in the particle, respectively. Note that we have selected the components $p_x$ and $m_z$ since they are the dominant dipolar responses, as shown in Fig.~\ref{fig_u_shape}c. The component $m_y$ is negligible here due to the small thickness of the particle. Then, integrating the Maxwell stress tensor on a sphere far away from the particle yields the time-averaged optical force (see the Appendix)

\begin{equation}
	\label{eq_fy}
	\begin{split}
			\langle\ve{F}\rangle &= \int_0^\pi\int_0^{2\pi} \langle\te{T}\rangle\cdot\ve{n} r^2\sin\theta~d\theta d\phi \\&= \ve{\hat{y}}A_\textrm{p}A_\textrm{m}\frac{k^3\mu_0\omega}{60\pi}\cos{\Delta\varphi},
	\end{split}
\end{equation}
where $\Delta\varphi = \varphi_\text{p}-\varphi_\text{m}$. This confirms that the combination of the dipolar responses $p_x$ and $m_z$ induces a lateral force in the $y$-direction and that the phase difference between these contributions is responsible for the sign of this force.~\cite{kiselev2020} Note that since we have not taken into account all multipolar contributions from the particle, and also ignored the incident field in~\eqref{eq_dip}, the longitudinal force does not appear in~\eqref{eq_fy}.

\begin{figure*}[b!]
	\centering
	\includegraphics[width=2\columnwidth]{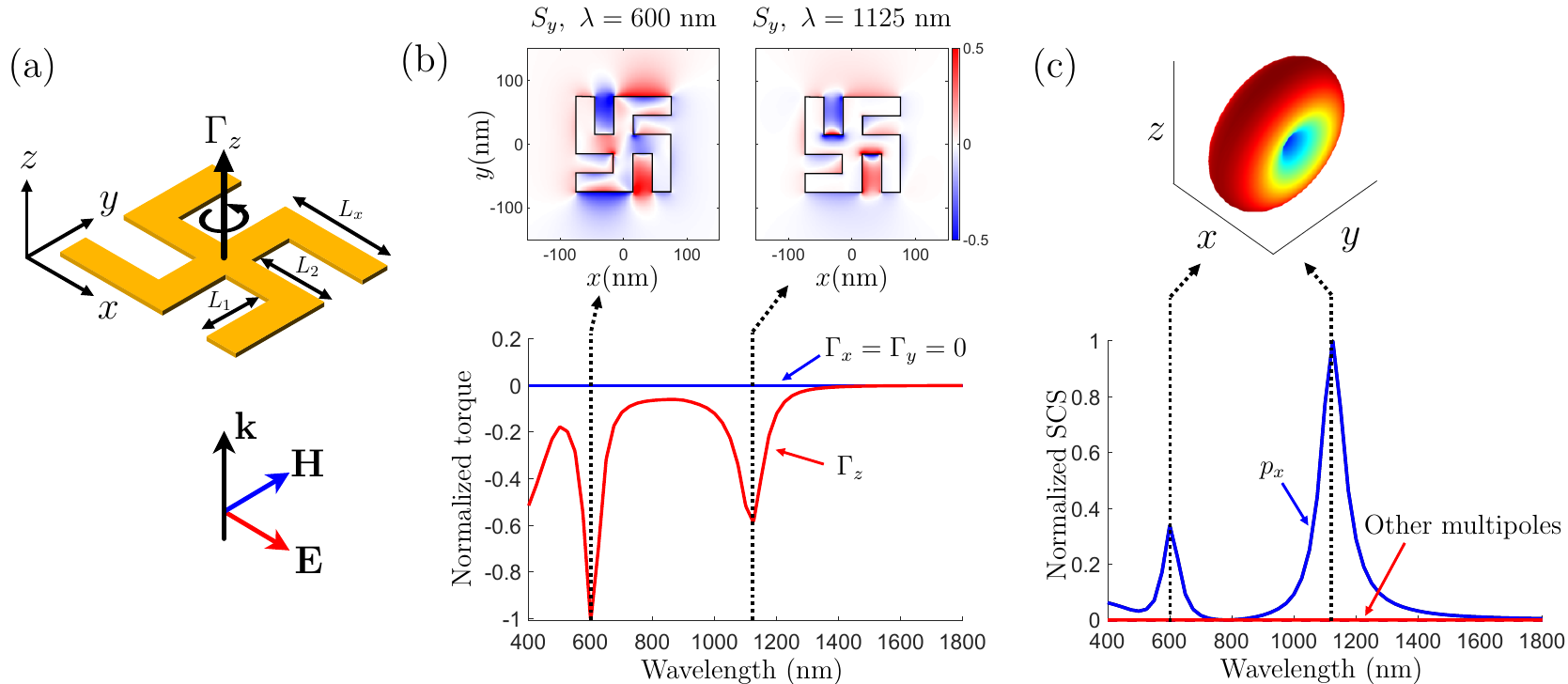}
%			\psfragfig*[width=2\columnwidth]{figs/gammadion}{
%				\psfrag{a}[][][1.]{(a)}
%				\psfrag{b}[][][1.]{(b)}
%				\psfrag{c}[][][1.]{(c)}
%				\psfrag{l2}[][][0.6]{$L_2$}
%				\psfrag{l1}[][][0.6]{$L_1$}
%				\psfrag{lx}[][][0.6]{$L_x$}
%				\psfrag{hh}[][][0.7]{$S_y,~\lambda=600$~nm}
%				\psfrag{ii}[][][0.7]{$S_y,~\lambda=1125$~nm}
%				\psfrag{aa}[][][0.7]{$p_x$}
%				\psfrag{bb}[][][0.7]{$m_y$}
%				\psfrag{cc}[][][0.7]{$p_x$}
%				\psfrag{dd}[][][0.7]{Other multipoles}
%				\psfrag{ff}[][][0.7]{$\Gamma_z$}
%				\psfrag{gg}[][][0.7]{$\Gamma_x=\Gamma_y=0$}
%				\psfrag{p}[][][0.7]{Normalized torque}
%				\psfrag{i}[][][0.7]{Wavelength (nm)}
%				\psfrag{j}[][][0.7]{Normalized SCS}
%				\psfrag{k}[][][1.]{$\ve{k}$}
%				\psfrag{e}[][][1.]{$\ve{E}$}
%				\psfrag{h}[][][1.]{$\ve{H}$}
%				\psfrag{t}[][][1.]{$\Gamma_z$}
%				\psfrag{xx}[][][0.7]{$x (\text{nm})$}
%				\psfrag{yy}[][][0.7]{$y (\text{nm})$}
%				\psfrag{x}[][][1.]{$x$}
%				\psfrag{y}[][][1.]{$y$}
%				\psfrag{z}[][][1.]{$z$}}
	\caption{Optical torque acting on a gold Gammadion particle. (a)~Illumination condition. (b)~Components of the optical torque and Poynting vector $S_y=\text{Re}\{E_zH_x^*-E_x^*H_z\}/2$. (c) Multipolar decomposition showing that only the $x$-component of the electric dipole $\ve{p}$ is dominant at both resonant frequencies. The particle dimensions are $L_x=90$~nm, $L_1=L_2=60$~nm and a thickness of 30~nm.}
	\label{fig_gamm}
\end{figure*}

We now understand that such a lateral force stems from the interference between multipolar contributions that are excited due to the asymmetry of the particle, which leads to asymmetric radiation~\cite{bliokh2015,achouri2020,kiselev2020}. This implies that the particle scatters more light either in the $+y$ or the $-y$ directions and is therefore subjected to a force by the action-reaction principle. This may be verified by considering that the optical force acting on a small particle illuminated by a plane wave may be expressed as proportional to the Poynting vector~\cite{chen2011a}. For illustration, we have plotted the time-averaged $y$-component of the Poynting vector, $S_y$, at the two resonant frequencies in Fig.~\ref{fig_u_shape}b, which are in good agreement with the corresponding far-field radiation patterns in Fig.~\ref{fig_u_shape}c since $S_y$ is mostly negative at $650$~nm and mostly positive at $1225$~nm.

\section{Optical Angular Momentum}
\label{sec_TAM}

Putting the pieces of the puzzle together, we now use four of the U-shaped particle in Fig.~\ref{fig_u_shape}a to form the Gammadion particle depicted in Fig.~\ref{fig_gamm}a and investigate the torque acting on it. Numerical simulations combined with the expression~\eqref{eq_T} for the torque using the Maxwell stress tensor given in the Appendix, reveal that the Gammadion particle is subjected to a $z$-oriented torque, $\Gamma_z$, that is thus collinear with the direction of the illumination, as plotted in Fig.~\ref{fig_gamm}b.

Now, remember that this particle is not geometrically chiral since it exhibits a reflection symmetry along the $z$-direction, it is therefore not obvious to understand why such a particle would rotate on itself when illuminated at normal incidence. If the particle was illuminated at oblique incidence, we could explain the presence of this longitudinal torque in terms of conventional (dipolar) pseudochirality (extrinsic chirality), which corresponds to chiral responses from an achiral object that depend upon the illumination condition~\cite{tretyakov2001,plum2009,plum2009a}. However, such effects are a priori not supposed to appear at normal incidence, although, as we shall see in the next section, that is valid if only dipolar responses are considered.

For the time being, we shall only restrict our analysis to the consideration of the torque within the action-reaction framework. Accordingly, we now investigate the distribution of the Poynting vector in the vicinity of the particle. As an illustration, the $y$-component of the Poynting vector is plotted in Fig.~\ref{fig_gamm}b at the two resonant frequencies. Such a peculiar distribution of power flow around the particle suggests the existence of angular momentum~\cite{liu2010b}. As a verification, we use the definition of the electromagnetic angular momentum given by~\cite{jackson1999}
\begin{equation}
	\label{eq_Lem}
	\ve{L}_\text{em} = \frac{1}{c^2}\int \ve{r}\times\ve{S} ~dV,
\end{equation}
where $\ve{r}$ is the position vector from the center of the particle, $\ve{S}$ is the Poynting vector and $c$ is the speed of light in vacuum. Applying~\eqref{eq_Lem} to the structure in Fig.~\ref{fig_gamm} leads to $\ve{L}_\text{em}\cdot\ve{\hat{z}} \neq 0$. This clearly means that the particle must be subjected to a torque, $\Gamma_z$. Indeed, since there is angular momentum in the field scattered by the particle but not in the field exciting it (a linearly polarized plane wave), the particle must therefore experience a torque by conservation of angular momentum.~\cite{liu2010b} While this explanation might be sufficient to understand the existence of a longitudinal optical torque, it does not provide a complete explanation of this phenomenon, as we shall see in the next section.

Interestingly, the response of the particle is this time totally dominated by an $x$-oriented dipolar response, as shown in Fig.~\ref{fig_gamm}c. This is consistent with the symmetrical nature of the particle and implies that the latter is not subjected to a lateral force since the radiation pattern is symmetric.

\section{Multipolar Pseudochirality}

To provide a deeper and more complete explanation for the presence of a longitudinal torque in such an achiral structure, we now investigate its multipolar responses. To do so, we shall concentrate on the particle dipolar and quadrupolar responses related to fields and field derivatives (nonlocal) excitations such that the relationship between responses and excitations may be given by~\cite{achouri2022}
\begin{equation}
	\label{eq_Qdip}
	\begin{bmatrix}
		p_i\\
		m_i\\
		Q_{il}\\
		S_{il}
	\end{bmatrix}\propto
	\begin{bmatrix}
		\alpha_{\text{ee}}^{ij} & \alpha_{\text{em}}^{ij} & \alpha_{\text{ee}}^{'ijk} & \alpha_{\text{em}}^{'ijk}\\
		\alpha_{\text{me}}^{ij} & \alpha_{\text{mm}}^{ij} & \alpha_{\text{me}}^{'ijk} & \alpha_{\text{mm}}^{'ijk}\\
		Q_{\text{ee}}^{ilj} & Q_{\text{em}}^{ilj} & Q_{\text{ee}}^{'iljk} & Q_{\text{em}}^{'iljk}\\
		S_{\text{me}}^{ilj} & S_{\text{mm}}^{ilj} & S_{\text{me}}^{'iljk} & S_{\text{mm}}^{'iljk}
	\end{bmatrix}\cdot
	\begin{bmatrix}
		E_{j}\\
		H_{j}\\
		\partial_k E_{j}\\
		\partial_k H_{j}
	\end{bmatrix},
\end{equation}
where $p_i$ and $m_i$ are the electric and magnetic dipolar responses whereas $Q_{il}$ and $S_{il}$ are the electric and magnetic quadrupolar responses, respectively. The terms inside the matrix in~\eqref{eq_Qdip} correspond to dipolar and quadrupolar polarizability tensors that model the effective response of a particle. Note that we consider that the multipolar decomposition is performed in spherical coordinates, which naturally leads to symmetric and traceless (irreducible) multipole moments~\cite{alaee2018,riccardi2022}. This, combined with multipolar reciprocity conditions~\cite{achouri2021c}, imposes several conditions on the polarizability tensors in~\eqref{eq_Qdip} that reduce the total number of independent polarizability components, as discussed in Refs.~\cite{achouri2022,tiukuvaara2023}. 

\begin{figure}[ht!]
	\centering
	\includegraphics[width=\columnwidth]{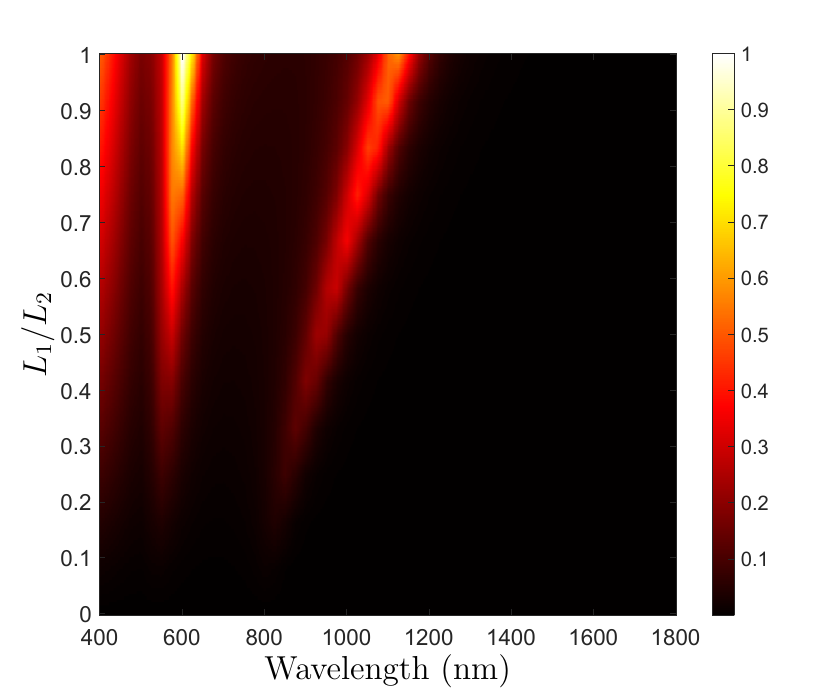}
	%			\psfragfig*[width=\columnwidth]{figs/torque}{
	%					\psfrag{x}[][][0.8]{Wavelength (nm)}
	%					\psfrag{y}[][][0.8]{$L_1/L_2$}}
	\caption{Normalized absolute value of the longitudinal torque, $\Gamma_z$, versus wavelength and gammadion shape asymmetry.}
	\label{fig_torque}
\end{figure}

Our goal here is not to develop a complete model that would describe the response of a given particle and compute all of its polarizabilities but, rather, to simply find out which components exist in~\eqref{eq_Qdip} and, from that, deduce the behavior of that particle. To do so, we only need to consider the particle spatial symmetries. Indeed, as we shall see next, the symmetries play a major role in the existence and strength of the torque acting on the particle.  As an illustration, consider the asymmetry ratio $L_1/L_2$, which equals 1 for the Gammadion particle in Fig.~\ref{fig_gamm}a and 0 for a simple symmetric square cross for which $L_1=0$ and $L_2=60$~nm. The evolution of the longitudinal torque, $\Gamma_z$, versus wavelength and the ratio $L_1/L_2$ is plotted in Fig.~\ref{fig_torque}. This simulation clearly shows that the torque is proportional to $L_1/L_2$ and thus disappears for a symmetric flat particle such as a square cross for which $L_1/L_2= 0$.

To connect the spatial symmetries of the particle to its material parameters, we refer to the Neumann's principle, which states that if a particle is invariant under certain spatial symmetries, then its effective material parameters should also be invariant under the same symmetry operations. This allows us to reduce the complexity of~\eqref{eq_Qdip} so that only the components that are relevant for the considered particle are left in the multipolar polarizability tensors. This may be achieved by recursively applying invariance conditions for each spatial symmetry operation, specified by the matrix $\te{\Lambda}$, to each of the material tensors in~\eqref{eq_Qdip} following the procedure described in Ref.~\cite{achouri2022} These invariance conditions are given, for an arbitrary tensor $\te{T}$ of order 2, 3 and 4, by~\cite{achouri2022}
\begin{subequations}
	\label{eq_InvCondT}
	\begin{align}
		T_{ij} &= a\Lambda_{im}\Lambda_{jk}T_{mk},\label{eq_InvCondT2}\\
		T_{ijk} &= a\Lambda_{il}\Lambda_{jm}\Lambda_{kn}T_{lmn},\\
		T_{ijkl} &= a\Lambda_{im}\Lambda_{jn}\Lambda_{ko}\Lambda_{lp}T_{mnop},
	\end{align}
\end{subequations}
where $a = \text{det}(\te{\Lambda})^n$ with $n=0$ for the `ee' or `mm' tensors in~\eqref{eq_Qdip} and $n=1$ for the `em' or `me' tensors, respectively.

\begin{figure*}[b!]
	\centering
	\includegraphics[width=2\columnwidth]{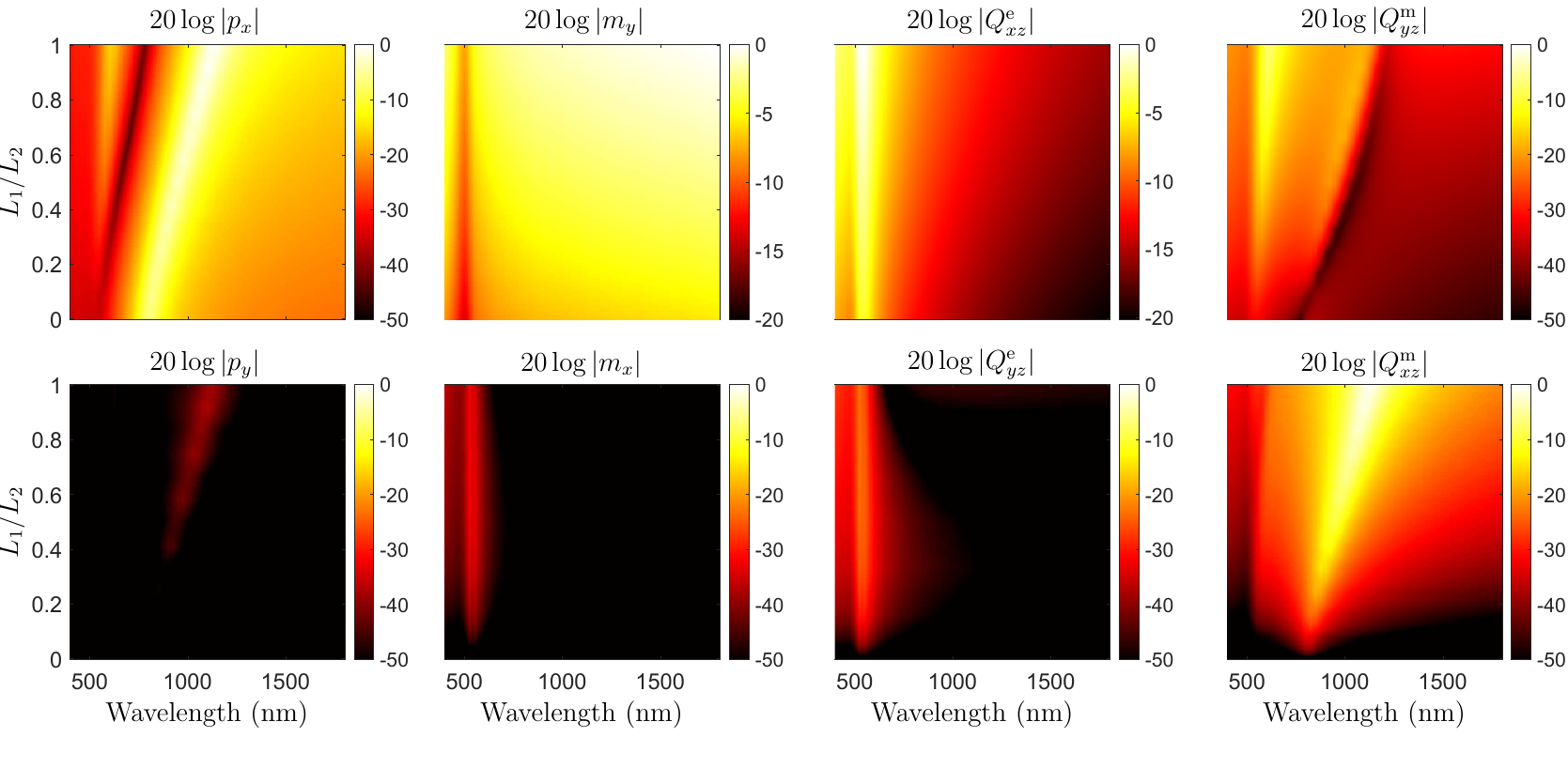}
	%		\psfragfig*[width=2\columnwidth]{figs/mult-gam}{
	%			\psfrag{a}[][][0.7]{$20\log|p_x|$}
	%			\psfrag{b}[][][0.7]{$20\log|m_y|$}
	%			\psfrag{c}[][][0.7]{$20\log|Q_{xz}|$}
	%			\psfrag{d}[][][0.7]{$20\log|S_{yz}|$}
	%			\psfrag{e}[][][0.7]{$20\log|p_y|$}
	%			\psfrag{f}[][][0.7]{$20\log|m_x|$}
	%			\psfrag{g}[][][0.7]{$20\log|Q_{yz}|$}
	%			\psfrag{h}[][][0.7]{$20\log|S_{xz}|$}
	%			\psfrag{x}[][][0.7]{Wavelength (nm)}
	%			\psfrag{y}[][][0.7]{$L_1/L_2$}}
	\caption{Normalized multipolar components versus wavelength and particle asymmetry ratio $L_1/L_2$. The top and bottom rows correspond to co-polarized and cross-polarized responses, respectively. The plots are normalized by dividing separately each column by the maximum value of either the co- or cross-polarized component.}
	\label{fig_mult_gam}
\end{figure*}

As a simple example, consider the U-shaped particle in Fig.~\ref{fig_u_shape}a. This particle is symmetric along the $z$ and $x$-axes and is thus invariant under the symmetry operations $\te{\Lambda}=\sigma_z$ and $\te{\Lambda}=\sigma_x$, where \mbox{$\sigma_n=\te{I}-2\ve{n}\ve{n}$} with $\ve{n}$ being the reflection axis and $\te{I}$ the identity matrix. Using~\eqref{eq_InvCondT2} successively\footnote{The approach to perform this operation is thoroughly described in Ref~\cite{achouri2022}.} with these two symmetry operations on the polarizability tensors $\te{\alpha}_\text{ee}$, $\te{\alpha}_\text{mm}$, $\te{\alpha}_\text{em}$ and $\te{\alpha}_\text{me}$ in~\eqref{eq_Qdip} may be used to identify nonzero polarizability tensor components that correspond to the prescribed symmetry operations. For instance, applying~\eqref{eq_InvCondT2} to the polarizability tensor $\te{\alpha}_\text{me}$ leads to the system
\begin{equation}
 \begin{cases}
      \sigma_x\cdot\te{\alpha}_\text{me}\cdot\sigma_x+\te{\alpha}_\text{me}=0,\\
     \sigma_z\cdot\te{\alpha}_\text{me}\cdot\sigma_z+\te{\alpha}_\text{me}=0,
    \end{cases}
\end{equation}
which when solved reveals that the component $\alpha_\text{me}^{zx}$ must be nonzero for the structure shown in Fig.~\ref{fig_u_shape}a. This leads to an effective excitation of the magnetic dipole along the $z$-axis since the bianisotropic polarizability $\alpha_\text{me}^{zx}$  connects the excitation $E_x$ to the response $m_z$ as $m_z \propto \alpha_\text{me}^{zx}E_x$.  This shows that the simple investigation of the particle symmetries allows us to predict its electromagnetic responses and, ultimately, the forces and torques acting on it under a given illumination condition.

Let us now apply this procedure to the Gammadion particle of Fig.~\ref{fig_gamm}a. This particle is invariant under $\te{\Lambda}=\sigma_z$ and $\te{\Lambda}=C_{4z}$, where $C_{4z}$ represents the $90^\circ$-rotation matrix along the $z$-axis. The application of~\eqref{eq_InvCondT} with these symmetry operations on the material parameters of~\eqref{eq_Qdip} reveals that $\te{\alpha}_\text{me} = \te{\alpha}_\text{em} = 0$ and that $\te{\alpha}_\text{ee}$ and $\te{\alpha}_\text{mm}$ are diagonal matrices. This counter-intuitive result means that the response of a Gammadion particle \emph{cannot} be distinguished from that of a flat symmetric particle, such as a square cross for which $L_1/L_2=0$, at least when considering only the conventional dipolar polarizabilities  $\te{\alpha}_\text{ee}$, $\te{\alpha}_\text{mm}$, $\te{\alpha}_\text{em}$ and $\te{\alpha}_\text{me}$.

However, the consideration of the symmetric properties of the higher-order polarizabilities reveals the presence of some nonzero off-diagonal terms for a Gammadion particle that are zero in the case of a square cross. In particular, it was recently shown that the components $\alpha_{\text{em}}^{'yzy}$, $\alpha_{\text{me}}^{'xzx}$, $\alpha_{\text{me}}^{'xzx}$ and $\alpha_{\text{em}}^{'yzy}$ are nonzero for a Gammadion particle, as reported in Ref~\cite{achouri2022}. This leads to an effective excitation of the following dipolar responses:
\begin{subequations}
	\label{eq_pm}
	\begin{align}
		p_y \propto \alpha_{\text{em}}^{'yzy} \partial_zH_y,\\
		m_x \propto \alpha_{\text{me}}^{'xzx} \partial_zE_x,
	\end{align}
\end{subequations}
as well as the following quadrupolar ones:
\begin{subequations}
	\label{eq_QS}
	\begin{align}
		Q_{yz} \propto Q_\text{em}^{yzy} H_y=-\alpha_{\text{me}}^{'yzy} H_y,\\
		S_{xz} \propto S_\text{me}^{xzx}E_x=-\alpha_{\text{em}}^{'xzx} E_x,
	\end{align}
\end{subequations}
%
%\footnote{Additionally, one can show a reciprocal connection between $\alpha_{\text{em}}^{'ijk}$ and  $Q_\text{em}^{ijk}$  and additionally between $\alpha_{\text{me}}^{'ijk}$ and $S_\text{me}^{ijk}$, see Ref~\cite{achouri2022}.}
where the reciprocity conditions have been applied to connect together the components of~\eqref{eq_pm} and~\eqref{eq_QS} following the procedure described in Refs~\cite{achouri2021c,achouri2022}. Considering that the illumination is $x$-polarized, it is clear that the multipolar responses in~\eqref{eq_pm} and~\eqref{eq_QS} correspond to \emph{cross-polarized} effects meaning that such a particle does indeed rotate the polarization of light. Since the particle is not geometrically chiral but still exhibits a chiral response due to multipolar contributions, we associate this effect with a form of multipolar pseudochirality (extrinsic chirality), which, in contrast to the previously reported (dipolar) extrinsic chiral effects~\cite{plum2009,plum2009a}, may take place even at normal incidence. Naturally, the particle also exhibits \emph{co-polarized} dipolar and quadrupolar responses, respectively given by
\begin{subequations}
	\label{eq_Cpm}
	\begin{align}
		p_x \propto \alpha_{\text{ee}}^{xx} E_x,\\
		m_y \propto \alpha_{\text{mm}}^{yy} H_y,
	\end{align}
\end{subequations}
and 
\begin{subequations}
	\label{eq_CQS}
	\begin{align}
		Q_{xz} \propto Q_{\text{ee}}^{'xzzx} \partial_zE_x,\\
		S_{yz} \propto S_{\text{mm}}^{'yzzy} \partial_zH_y.
	\end{align}
\end{subequations}
It should be noted that the co-polarized dipolar responses in~\eqref{eq_Cpm} are directly proportional to the fields, whereas the cross-polarized ones in~\eqref{eq_pm} are proportional to the derivative of the fields in the $z$-direction. This indicates that the cross-polarized responses stem from longitudinal nonlocal (spatially dispersive) effects thus suggesting that the thickness of the particle compared to the wavelength plays a role in the strength of these responses.

To verify that the multipoles in~\eqref{eq_pm} to~\eqref{eq_CQS} are indeed excited, we have performed a multipolar decomposition versus wavelength and asymmetry ratio $L_1/L_2$ whose results are plotted in Fig.~\ref{fig_mult_gam}. By inspecting the dependencies of these plots along the wavelength, we can see that the co-polarized responses (top row) always exist irrespectively of the ratio $L_1/L_2$. However, the cross-polarized responses (bottom row) progressively vanish as $L_1/L_2$ tends to 0, which confirms that the asymmetry of the particle plays an essential role in their existence. 

Now, to demonstrate that the existence of these cross-polarized multipolar responses is responsible for the observed longitudinal torque, we consider the formula derived in Ref.~\cite{vesperinas2015} and that expresses the electromagnetic torque in terms of dipolar responses as
\begin{equation}
\label{eq_vespTorque}
	\langle \ve{\Gamma}\rangle = -\frac{k^3}{3}\text{Im}\left\{\frac{1}{\epsilon}\ve{p}^*\times\ve{p}+\mu\ve{m}^*\times\ve{m}\right\},
\end{equation}
where $k$, $\epsilon$ and $\mu$ are the wavenumber, the permittivity and permeability of the background medium, respectively. It becomes clear from~\eqref{eq_vespTorque} that the existence of a torque along the $z$-direction necessarily requires the simultaneous excitation of co- and cross-polarized electric ($p_x$ and $p_y$) and/or magnetic ($m_x$ and $m_y$) dipoles. By substituting the value of the simulated dipole moments shown in Fig.~\ref{fig_mult_gam} into~\eqref{eq_vespTorque}, we can completely retrieve the torque that was previously found and plotted in Figs.~\ref{fig_torque} and~\ref{fig_gamm}b. This confirms that the longitudinal torque, $\Gamma_z$, stems from multipolar pseudochiral responses. \\

At this point, one might be faced with a paradox when comparing the results from Sec.~\ref{sec_TAM} and that of the current section. Indeed, on one hand, the existence of a longitudinal torque was explained in Sec.~\ref{sec_TAM} by the presence of a power flow distribution around the particle that leads to angular momentum. Looking at the Poynting vector field in Fig.~\ref{fig_gamm}b, one might a priori associate such a distribution with \emph{orbital} angular momentum, as was for instance done in Ref~\cite{liu2010b}. On the other hand, we have seen in the current section that it is the excitation of cross-polarized responses that leads to such a torque, which may intuitively be associated with \emph{spin} angular momentum since rotation of polarization is involved. To resolve this paradox, consider the derivation of~\eqref{eq_vespTorque}, which is well detailed in Ref.~\cite{vesperinas2015}. This formula corresponds to the total torque acting on the particle and is in fact composed in \emph{equal} parts of both spin and orbital angular momentum. Therefore, since~\eqref{eq_vespTorque} combined with the data in Fig.~\ref{fig_mult_gam} is in perfect agreement with the computed torque in Fig.~\ref{fig_torque}, we conclude that this torque is equally composed of spin and angular orbital momenta.

\section{Conclusion}

We have shown that achiral nano-particles may be subjected to an optical torque even when illuminated at normal incidence. It was demonstrated that such particles do indeed exhibit chiral responses that stem from multipolar bianisotropic effects that correspond to a form of pseudochirality. For nano-particles, such multipolar pseudochiral responses, although weak, are sufficient to induce mechanical rotation by conservation of angular momentum. This work also shows that, in this case, the optical torque is directly related to the thickness and the asymmetry of the particle, and may therefore serve as a foundation for the development of highly efficient nano-rotors.

\section*{Acknowledgements}

We gratefully acknowledge funding from the European Research Council (ERC-2015-AdG-695206 Nanofactory) and from the Swiss National Science Foundation (project PZ00P2\_193221).

\section{Appendix}

\subsection{Electromagnetic Torque computation}

The time-averaged Maxwell stress tensor for an object surrounded by vacuum reads~\cite{jackson1999}
\begin{equation}
\begin{aligned}
	\label{eq_MST}
	\langle\te{T}\rangle &= \frac{1}{2}\textrm{Re}\Bigr[\epsilon_0\ve{E}\ve{E}^*+\mu_0\ve{H}\ve{H}^* \\
\qquad &-\frac{1}{2}\left(\epsilon_0|\ve{E}|^2+\mu_0|\ve{H}|^2\right)\te{I}\Bigr]
\end{aligned}
\end{equation}
where $\te{I}$ is the identity matrix. It follows that the time-averaged electromagnetic force acting on the object may be obtained by integrating~\eqref{eq_MST} over a sphere $S$ surrounding the object as
\begin{equation}
	\langle\ve{F}\rangle = \int_S \langle\te{T}\rangle\cdot\ve{n}~dS,
\end{equation}
whereas the corresponding torque is found using
\begin{equation}
	\label{eq_T}
	\langle\ve{\Gamma}\rangle = \int_S \ve{r}\times\left(\langle\te{T}\rangle\cdot\ve{n}\right)~dS.
\end{equation}

\bibliography{All.bib}
%\bibliography{../../../../References/All.bib}
	
\end{document}